\documentclass[11pt, authoryear]{llncs}
\usepackage{booktabs} 
\usepackage[ruled]{algorithm2e} 

\SetAlFnt{\small}
\SetAlCapFnt{\small}
\SetAlCapNameFnt{\small}
\SetAlCapHSkip{0pt}
\IncMargin{-\parindent}

\usepackage{graphicx}
\usepackage{natbib}
\usepackage{amsfonts, amsmath}
\usepackage{bm}
\usepackage{bbm}
\usepackage{booktabs}
\usepackage{mathtools}
\usepackage{fullpage}
\usepackage{enumerate}
\DeclareMathOperator\sign{sign}
\DeclareMathOperator\supp{supp}
\DeclareMathOperator\TR{TR}
\DeclareMathOperator\LTR{LTR}
\DeclareMathOperator\CTR{CTR}
\DeclareMathOperator\TRG{TRG}
\DeclareMathOperator\COL{COL}
\DeclareMathOperator\BDG{BDG}
\DeclareMathOperator\REI{REI}
\DeclareMathOperator\LOW{LOW}
\DeclareMathOperator\HIGH{HIGH}
\DeclareMathOperator\NEXT{NEXT}
\DeclareMathOperator\HIGHER{HIGHER}
\DeclareMathOperator\RANGE{RANGE}
\DeclareMathOperator\RES{RES}
\newcommand{\erf}{\text{erf}}
\newcommand{\E}{\mathop{\mathbb E}}

\newcommand{\n}{\normalsize}
\newcommand{\s}{\normalsize}

\title{Budget-Constrained Reinforcement of Ranked Objects}

\author{Written by Amir Ban\inst{1} and Moshe Tennenholtz\inst{2}\
\institute{Weizmann Institute of Science, Rehovot, Israel  \email{amirban@netvision.net.il} \and Faculty of Industrial Engineering and Management, Technion, Israel Institute of Technology  \email{moshet@ie.technion.ac.il}}}

\begin{document}

\maketitle

\begin{abstract}
Commercial entries, such as hotels, are ranked according to score by a search engine or recommendation system, and the score of each can be improved upon by making a targeted investment, e.g., advertising. We study the problem of how a principal, who owns or supports a set of entries, can optimally allocate a budget to maximize their ranking. Representing the set of ranked scores as a probability distribution over scores, we treat this question as a game between distributions.

We show that, in the general case, the best ranking is achieved by equalizing the scores of several disjoint score ranges. We show that there is a unique optimal reinforcement strategy, and provide an efficient algorithm implementing it.
\end{abstract}

\section{Introduction}

Suppose you have a YouTube channel, and you want to promote its videos. To kick-start your channel's popularity, you want to improve YouTube's ranking of your videos, and you have a budget to spend on that. Which of your videos should you spend on, and in what proportion? You can augment the view count of each video by targeted advertising. You have 5 videos, currently with 6K, 3K, 700, 500 and 100 views, and your budget is sufficient to add 10K views. How do you spend the budget to get the best average rank for your videos?

Clearly this depends on the distribution of view count among YouTube videos. You don't know the distribution, but you reasonably assume that their frequency decreases with higher view count. Under this assumption, we will show (see Example \ref{decreasing} below) that your best course is to add $0, 575, 2875, 3075$ and $3475$ views, respectively, for your 5 videos. This will have the effect of increasing the view count of your 4 lowest-ranked videos to 3575.

If instead of videos, you wish to promote 5 hotels on a recommendation system, and there are 85 hotels in total in your area, you might take a less statistical point-of-view. Assuming you have a budget sufficient to generate $x$ additional recommendations,
how do you divide $x$ between your 5 hotels given the exact current status of all area hotels on the recommendation system?

The above are two examples of the problem of budget-constrained optimization of the ranking of entries supported by a principal in a ranked pool of entries, the rationale for which is that the higher an entry's rank, the more likely it is to be visible in search results. Entries not supported by the principal are called the {\em complement}, and are a static and passive feature of the problem. Each entry has a numerical score, and the ranking is by highest to lowest. Suppose your budget enables you to increase the score of any entry by $1$. This may improve the rank of one entry by dozens of places, while another's, not at all. However, if the budget is doubled, it may be the latter entry whose rank will gain more than the former. So a greedy solution is not the right one for this problem. 

We detail a full solution to this problem, providing a characterization of the optimal reinforcement strategy, as well as algorithms to construct it. Treating the score of a ranked entry as a random variable, its distribution fully describes the scores of a set of entries. Our optimal reinforcement problem then becomes a question of finding a distribution $\bm{A}$ that is a best response to a given complement distribution $\bm{c}$, according to some utility function defined on the two distributions. The initial supported entry scores constrain the solution distribution $\bm{A}$, that it must (first-order) stochastically dominate an initial distribution $\bm{a}$. The budget constraint translates to a limit on the difference in expectations between $\bm{A}$ and the initial distribution $\bm{a}$.

Referring to the cumulative distribution functions (c.d.f.) of the relevant distributions, our characterization shows that an optimal reinforcement strategy must meet some requirements on the chords and tangents of these c.d.f. plots. We use the characterization to derive a constructive reinforcement algorithm that is optimal or $\epsilon$-optimal, depending on how the question is posed. The running time of the algorithm is almost-surely linear in the number of ranked entries.

The optimal reinforcement strategy, as it turns out, identifies one or more {\em target scores}, and correspondingly, a set of disjoint segments of scores (i.e., positive reals) each of whose high endpoint is a target score. Every supported entry whose score falls in one of these disjoint segments is reinforced to its respective target score. An exceptional case is where two or more target scores meet a condition we call collinearity, when entries may be reinforced to any score between the collinear targets. The simplest mode of our solution is for continuous unimodal distributions, of which our opening paragraphs provide an example. We devote a section to this commonly-occurring case.

Our contribution is in the complete solution of this problem, which, to our knowledge, has not been previously attempted. Also, a linear-time solution is probably unanticipated, and may raise interest in applying it on large-scale data. 

\subsection{Related Literature}

Ranking systems, and in particular their capability for manipulation, have been extensively studied \cite{cheng2005sybilproof}, \cite{altman2006quantifying}, \cite{altman2007incentive}, \cite{altman2010axiomatic}. Getting optimal influence for a given budget, or vice versa, was considered by \cite{feldman2007budget} and \cite{alon2012optimizing}, and in bidding for advertising, by \cite{wu2018budget}. Manipulating elections is a closely-related subject, e.g. \cite{conitzer2007elections}.

Our original motivation was to investigate competitive reinforcement games, in which two principals have separate budgets to support a set of ranked entries.  What is sought are reinforcement
strategies in equilibrium. Stated thus, the problem is related to Colonel Blotto games, due to \cite{borel1953theory}, in which two colonels partition their men into detachments, assigned to a set of contested hills, each won by the colonel who assigned more men to that hill. 
An even better fit are General Lotto games \cite{hart2008discrete}, where after the partition by each side into detachments, one detachment is chosen at random from each army, and the bigger one wins.

A solution of such a problem, with asymmetric budgets, is given in \cite{sahuguet2006campaign}, where it is framed as an allocation problem in competitive campaign spending. \cite{hart2008discrete} 
represents the partition of resources as a probability distribution, demonstrating that this reduces the problem to a game between probability distributions.

While studying this background, we realized that the non-competitive reinforcement problem is far from trivial; it is more elementary and of more practical value than the competitive one, and to the best of our knowledge it has not been addressed in the literature. 

The rest of this paper is organized as follows. Section~\ref{the-problem} describes the reinforcement problem. Section~\ref{algorithm} details our method and algorithms and provides a numerical example. In Section~\ref{analysis} we characterize the optimal solution of the problem, and prove the correctness of the algorithms given.
In Section~\ref{discussion} we offer concluding remarks.

\section{The Problem}
\label{the-problem}

There is a set of entries $\mathcal{W}$, each attached with a positive numerical score, and ranked by score, e.g., by a search engine or recommendation system. A subset of those entries $\mathcal{A}$ are supported by a principal who is endowed with a budget, defined as the {\em total} score improvement of all supported entries. The principal partitions this budget among his supported entries, and wishes to do so in a way that maximizes the {\em average rank} of his supported entries\footnote{Improving the average rank by deleting an entry is not intended. To disincentivise that, assume that a deleted entry has infinite rank.}. The budget is $|\mathcal{A}| p_A$, where $p_A > 0$ is the average budget per entry.

Ties in scores are broken in favour of the principal, i.e. in favour of entries in $\mathcal{A}$. Ties between entries in $\mathcal{A}$ are broken in some arbitrary, but fixed manner. Our choice of tie-breaking rule does not affect the solution, because, whatever rule is adopted, an arbitrarily small additional budget is needed to decide all ties in favour of the principal.

Define the {\em complement} set $\mathcal{C} := \mathcal{W} \setminus \mathcal{A}$. A measure of the average rank, represented as a number in $[0,1]$, and increasing with higher average rank, may be calculated by summing the signs of the differences of each element of $\mathcal{A}$ with each element in $\mathcal{C}$\footnote{Note that the ordinal rank of an element $x$ in $W$ is given by $\frac{1}{2}\{1 + |\mathcal{W}| - \sum_{y \in \mathcal{W}} \sign(x -  y)\}$, and $\sum_{y \in \mathcal{W}} \sign(x -  y)$ increases the higher $x$'s rank . Since $\sum_{x \in \mathcal{A}}\sum_{y \in \mathcal{A}} \sign(x - y) = 0$, it is necessary to count only the signs of the differences of a set with its complement ($\mathcal{C}$).}, i.e., marking the score of entry $x$  by $r_x$
\s\begin{align*}
u(\mathcal{A}, \mathcal{C}) := \frac{1}{|\mathcal{A}| |\mathcal{C}|}\sum_{x \in \mathcal{A}} \sum_{y \in \mathcal{C}} \sign(r_x - r_y)
\end{align*}\n

In many practical settings of the problem, e.g., videos in YouTube, where a channel owner wants to advance his videos, the cardinality of $\mathcal{C}$ is much larger than the cardinality of the
supported set $\mathcal{A}$. We shall consider the problem when
\begin{itemize}
\item \textbf{Exact version:} The set $\mathcal{C}$ and the scores of each of its elements is specified in full, and a reinforcement strategy considering this full information is sought.
\item \textbf{Statistical version:} The set $\mathcal{C}$ is known statistically, i.e., its scores are considered to be i.i.d. samples from some known distribution. This distribution is positive, with piecewise-differentiable density, and is specified analytically, in closed-form formulas.
\end{itemize}

Both versions have their merits: Assuming there is full information on the complement, the exact problem solves the actual question at hand. On the other hand, full information may not be available, or may not be practical to process. Even when it is, it may be just a short-lived snapshot of a fluid reality. Modelling the complement statistically is then a more robust solution.

In our problem, entries in each set are indistinguishable (the scores of two different entries can be exchanged without consequence). We will, from now on, treat the score of entry sets as probability distributions, as follows: Let $r_z$ denote the score of entry $z$ before reinforcement, and $R_z$ after reinforcement. Then entry set $\mathcal{A}$, before and after reinforcement, are described by positive random variables $\bm{a}$ and $\bm{A}$, respectively, whose c.d.f. is
\s\begin{align}
\label{F_a}
F_{\bm{a}}(x) = \frac{1} {|\mathcal{A}|} |\{z \in \mathcal{A}: r_z \leq x\}|\\
\label{F_A}
F_{\bm{A}}(x) = \frac{1}{|\mathcal{A}|} |\{z \in \mathcal{A}: R_z \leq x\}|
\end{align}\n

These distributions are not only discrete, but their densities are restricted to fractions with a given denominator ($|\mathcal{A}|$), since they represent an integral number of entries. We call such distributions {\em integral}.

On the other hand, the complement set $\mathcal{C}$ is described by a positive random variable $\bm{c}$ with a piecewise twice-differentiable c.d.f., which is equaled (in the exact problem) or approximated (in the statistical problem) by the complement distribution
\s\begin{align*}
\label{F_c}
F_{\bm{c}}(x) \simeq \frac{1}{|\mathcal{C}|} |\{z \in \mathcal{C} : r_z \leq x\}| \\
\end{align*}\n

The p.d.f. of $\bm{a}$ is the sum of Dirac delta functions
\s\begin{align*}
f_{\bm{a}}(x) &= \frac{1}{|\mathcal{A}|} \sum_{z \in \mathcal{A}} \mathbbm{1}_{r_z}(x)
\end{align*}\n

So the expectation of $\bm{a}$ is
\s\begin{align*}
\E[\bm{a}] = \int_0^\infty x f_{\bm{a}}(x) dx =  \int_0^\infty [1 - F_{\bm{a}}(x)] dx =  \frac{1}{|\mathcal{A}|} \sum_{z \in \mathcal{A}} r_z
\end{align*}\n

\noindent i.e., $\bm{a}$'s expectation is the average score before reinforcement. Similarly, $\bm{A}$'s expectation is the average score {\em after} reinforcement.
\s\begin{align*}
\E[\bm{A}] = \int_0^\infty x f_{\bm{A}}(x) dx =  \int_0^\infty [1 - F_{\bm{A}}(x)] dx =  \frac{1}{|\mathcal{A}|} \sum_{z \in \mathcal{A}} R_z
\end{align*}\n

Thus,
\s\begin{align}
\E[\bm{A} - \bm{a}] = \E[\bm{A}] - \E[\bm{a}] = p_A
\end{align}\n

Since $R_z \geq r_z$, $R_z \leq x \Rightarrow r_z \leq x$, and so, for every $x$, $F_{\bm{a}}(x) \geq F_{\bm{A}}(x)$. In other words, $\bm{A}$ (first-order) stochastically dominates $\bm{a}$.

\begin{figure}[tbp]
\centering
\includegraphics[height=0.24\textheight]{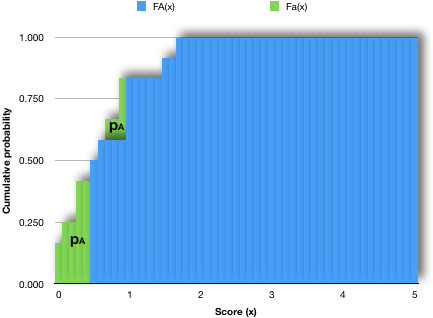}
\caption{Base $F_{\bm{a}}(x)$ and reinforced $F_{\bm{A}}(x)$ c.d.f.'s. The total difference area is budget $p_A$, with two reinforced segments: Entries with scores $0-0.5$ were reinforced to $0.5$, and with scores $0.7-1$, to $1$.}
\label{reinforced}
\end{figure}

Figure \ref{reinforced} exemplifies a base score distribution (in green, partly occluded by blue), stochastically dominated by a reinforced distribution (in blue). The area between the distributions is the budget $p_A$.

The problem can then be restated in terms of distributions.

\begin{definition} [Utility]
Define, for any random variables $X, Y$ \s$$U(X, Y) := \Pr[X \geq Y] - \Pr[X < Y]$$\n
\end{definition}

Then, for any random variables $X, Y$, we have
\s\begin{align*}
U(X, Y) &= \Pr[X \geq Y] - \Pr[X < Y] = 2\Pr[X \geq Y] - 1\nonumber \\
&= \int\limits_0^\infty \Pr[X = y]\big\{2\Pr[y \geq Y] - 1\big\}dy  \\
&=  \int\limits_0^\infty f_X(y) \big\{2F_Y(y) - 1\big\} dy
\end{align*}\n

\begin{problem} [Reinforcement Problem]
\label{single-principal}
Given an integral base distribution of principal $\bm{a}$, distribution of complement $\bm{c}$, and budget $p_A$, find an integral reinforced distribution $\bm{A}$ that maximizes
\s\begin{align}
U(\bm{A}, \bm{c}) \label{U_c} &=  \int\limits_0^\infty f_{\bm{A}}(y) \big[2F_{\bm{c}}(y) - 1\big] dy
\end{align}\n

\noindent subject to
\begin{itemize}
\item $\bm{A}$ stochastically dominates $\bm{a}$.
\item $\E[\bm{A} - \bm{a}] \leq p_A$.
\end{itemize}
\end{problem}

\section{Method}
\label{algorithm}

In this section, we describe a method to solve the {\em reinforcement problem} (Problem \ref{single-principal}). A later section will be devoted to proving the correctness of the method.

\subsection{Synopsis}

The method is based on the following idea. The optimal solution is to implement all possible reinforcements where the complement's average density between initial and final score exceeds a threshold, called $\alpha$ in the algorithms. Plotted on the complement's c.d.f., this means all reinforcements are made across chords steeper than $\alpha$. In practice, we select scores for which the complement's p.d.f. is $\alpha$, calling them target scores, and each is the endpoint of a segment of scores whose chord's gradient to the target score is at least $\alpha$. All supported entries (in $\mathcal{A}$) whose score falls in any such segment is reinforced to meet the corresponding segment end/target score.

All this is implemented by the {\em Basic Algorithm} (Algorithm \ref{basic} below) for a given parameter $\alpha$. The algorithm has no control over the budget it uses, which is a function $\BDG_\alpha$ of its parameter. However, as will be proved later (Proposition \ref{p_alpha}), this budget is non-increasing in $\alpha$.

Harnessing this monotonicity, the {\em Iterative Algorithm} (Algorithm \ref{supplementary} below), iteratively applies Algorithm \ref{basic} in a binary search to find an $\alpha$ parameter for which $\BDG_\alpha$ most closely approaches the given budget without exceeding it. Based on this, it constructs the sought-after optimal solution.

The algorithm is equally applicable to exact (list of values) and statistical (analytical formula) representations of the complement. For the latter, implementing the algorithm may need unspecified analytical or numerical steps (e.g., finding solutions of $f_{\bm{c}}(x) = \alpha$).

\subsection{A Numerical Example}
\label{numerical}

\begin{table}
  \begin{center}
    \caption{Example ranked scores. Supported entries are highlighted}
    \label{tab:table1}
    \begin{tabular}{c|r|r} 
      \textbf{Score} $x$ & $F_{\bm{a}}(x)$ & $F_{\bm{c}}(x)$\\
      \hline
      10 &&  0.125 \\
      {\em 10} & 0.25 & \\
      {\em 15} & 0.5 & \\
      24 && 0.25\\
      35 && 0.375\\
      {\em 40} & 0.75 & \\
      60 && 0.5 \\
      80 && 0.625 \\
      100 && 0.75 \\
      {\em 114} & 1 & \\
      200 && 0.875 \\
      220 && 1
    \end{tabular}
  \end{center}
\end{table}

\begin{figure}[tbp]
\centering
\includegraphics[height=0.5\textheight]{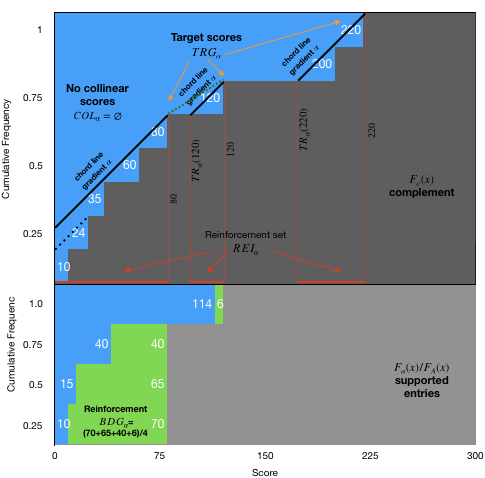}
\caption{Solution with Basic Algorithm for Gradient $\alpha$}
\label{solution}
\end{figure}

Consider the set of entries detailed in Table \ref{tab:table1}. There are $12$ entries, ranked by score (1st column), $4$ of which are supported, and $8$ in the complement. The c.d.f.'s of the supported and complement entries are shown in the 2nd and 3rd columns, respectively. See in Figure \ref{solution} a plot of the two c.d.f.'s. As the distributions are discrete, the c.d.f.'s are step functions. The top part of Figure \ref{solution} plots the complement's c.d.f., in dark grey. The bottom part plots the supported entries' c.d.f. in grey and green. The $x$-axis (scores) is common to both c.d.f.'s.

The figure shows the optimal reinforcement when the total budget $|\mathcal{A}| p_A$ equals $181$.

We draw parallel chord lines, whose common gradient is a given $\alpha$, from the {\em top} of each step in the complement's c.d.f.. Each chord line extends to the left, up to where it intersects with the c.d.f. plot, or until it intersects with the $y$-axis, whichever comes first. The score ($x$ value) of this intersection is called the {\em trace} of the chord line. E.g., the vertical dotted red line labeled $\TR_\alpha(220)$ in the figure is the trace of complement score $220$.

The set of scores that lie between complement scores and their traces is called the {\em reinforcement set}. It is marked in Figure \ref{solution} in three red segments on the $x$-axis. Note that only three complement scores contribute to the reinforcement set: $80, 120$ and $220$. The chord line of complement score $24$ (marked by a dotted black line), for example, does not contribute to the reinforcement set, because it is wholly contained in the contribution of complement score $80$. The same is true for the rest of the chord lines (not shown).

The 3 complement scores who do contribute to the reinforcement set are the {\em target scores}. Every supported score who is reinforced by the method, is reinforced to one of the target scores, and specifically to the next-higher target score.

The supported scores that are reinforced are those whose score is in the reinforcement set. Moving our attention in Figure \ref{solution} to the bottom plot, we see that all $4$ supported entries are in the reinforcement set, a coincidence. The green plot shows the reinforcement, with the number near the green/grey boundary showing the reinforcement for each supported entry. The total reinforcement, i.e., budget, is $70 + 65 + 40 + 6 = 181$, as required. Note that the budget is proportional to the green area.

However, if we are given a target budget, we have no direct method of determining which chord line gradient $\alpha$ results in such a budget. We have to search for it, and we do so using binary search, as described in the Iterative Algorithm.

\begin{figure}[tbp]
\centering
\includegraphics[height=0.5\textheight]{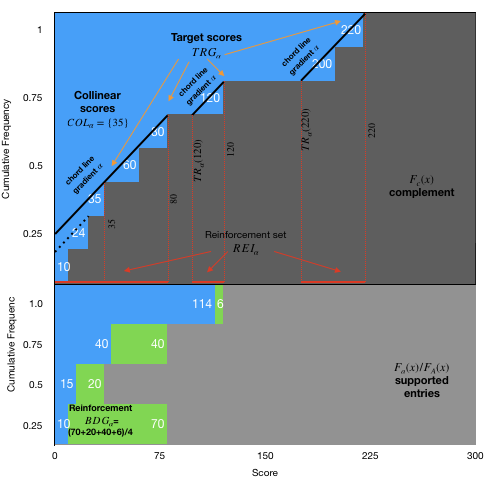}
\caption{Solution with Collinear Scores ($35$ and $80$)}
\label{collinear}
\end{figure}

A complicating, but unavoidable factor is the possibility of collinearity, where a score lies exactly on another score's chord line (for some gradient $\alpha$). In Figure \ref{collinear} we see a picture that is almost identical to that of Figure \ref{solution}, except that the gradient $\alpha$ is steeper just enough to make score $80$'s chord line meet the complement's c.d.f. at the top of the step for score $35$, with the result that the chord lines for complement scores $80$ and $35$ are collinear.

As in Figure \ref{solution}, all scores from $0$ to $80$ are in the reinforcement set. Supported scores $10, 15$ and $40$ should therefore be reinforced to the next-higher target score. But for the scores below $35$ ($10$ and $15$), which is the next-higher target score, $35$ or $80$? The answer given by our analysis is: either. In Figure \ref{collinear} score $10$ is reinforced to $80$ and score $15$ is reinforced to $35$, for a total budget of $136$. An equally optimal solution with the same budget is to promote score $10$ to $35$ and score $15$ to $80$. Since these two scores can be reinforced to either $35$ or $80$, this collinear gradient $\alpha$ is responsible for the optimal solution for budgets anywhere from $25 + 20 + 40 + 6 = 91$ to $70 + 65 + 40 + 6 = 181$. Given a budget of, e.g., $150$, our optimal solution would be as in Figure \ref{collinear}, leaving $150 - 136 = 14$ unused.

The reader will notice that the upper bound of this range is $181$, which is the same as the Figure \ref{solution} budget. This is because the gradient $\alpha$ of Figure \ref{solution} can be perturbed without change: So long as the chord lines do not touch a new score, the solution, and so the budget is unchanged.

\subsection{Preliminaries}


\begin{definition}[Target scores, supports and gradients]
A {\em target score} is a score $x$ for which $f_{\bm{a}}(x) < f_{\bm{A}}(x)$.

For a random variable $X$, $\supp(X)$ denotes its support, the set of values where its density is non-zero.

Define the {\em $x,y$-gradient} of $\bm{c}$ as \s$$C_x^y := \frac{F_{\bm{c}}(x) - F_{\bm{c}}(y )}{x - y}$$\n for $x \neq y$. $C_x^y$ is the slope of a chord of $F_{\bm{c}}(\cdot)$ between scores $x$ and $y$, and is, by definition, the average density between these two scores.
%
\end{definition}

If $\bm{c}$ has no atom at $x$, then for $\epsilon > 0$, $\lim_{\epsilon \to 0} C_x^{x+\epsilon}$ and $\lim_{\epsilon \to 0} C_x^{x-\epsilon}$ both exist, and are called, respectively, the left and right density of $\bm{c}$ at $x$. If $\bm{c}$'s c.d.f. is differentiable at $x$, they are equal. \s$$\lim_{\epsilon \to 0} C_x^{x+\epsilon} = \lim_{\epsilon \to 0} C_x^{x-\epsilon} = f_{\bm{c}}(x)$$\n


\begin{definition} [Traces]
\label{traces}
Given a gradient $\alpha$, a score $x$, and a complement c.d.f. $F_{\bm{c}}$, the {\em trace} of $x$ with $\alpha$, marked $\TR_\alpha(x)$, is the maximal score $z \in (0,x)$ for which $F_{\bm{c}}(x) + \alpha (z - x) = F_{\bm{c}}(z)$, or is $0$ if there is no such $z$. The line $F_{\bm{c}}(x) + \alpha (z - x )$ is called $x$'s {\em chord line} (for $\alpha$). In other words, $\TR_\alpha(x)$ is the $x$-value where $x$'s chord line, extended to the left of $x$, meets the plot of the complement's c.d.f., or meets the $y$-axis, whichever occurs first. See in Figures \ref{solution} and \ref{collinear}, e.g., the dotted red line labeled $\TR_\alpha(220)$.

In cases where $F_{\bm{c}}$ coincides with the chord line at $x$ for a left-neighborhood of $x$ (i.e., $F_{\bm{c}}(x - \epsilon) = F_{\bm{c}}(x) - \epsilon \alpha$ for all sufficiently small $\epsilon > 0$), we define $\TR_\alpha(x) = x$.

The segment $[\TR_\alpha(x), x)$ is called $x$'s $\alpha$-segment. See, e.g., the 3 red segments on the $x$-axis in Figure \ref{solution}.
\end{definition}


\begin{definition}[Segments]
\label{segments}
Let $\mathcal{Z} \subseteq \mathbb{R}_{>0}$ be a set of disjoint segments of scores, each of which may be open or closed at either end. Let $R \subseteq \mathcal{Z}$ be such a segment.  Define $\LOW(R)$ and $\HIGH(R)$ as its low and high, respectively, endpoint. Furthermore, for every score $x \in R$, define $\HIGH_\mathcal{Z}(x) := \HIGH(R)$ and $\LOW_\mathcal{Z}(x) := \LOW(R)$.
\end{definition}

\subsection{Basic Algorithm Notes}
\label{annotations}

The basic algorithm is detailed in Algorithm \ref{basic}. Following are notes referred to in the algorithm.

\begin{algorithm}[tb]
Given an integral distribution $\bm{a}$, a complement distribution $\bm{c}$, and a gradient $\alpha > 0$ (See Note \ref{alpha}), the algorithm outputs
\begin{itemize}
\item A reinforced integral distribution $\bm{A}$.
\item A positive real budget $\BDG_\alpha$.
\item A set of scores $\COL_\alpha$ (the {\em collinear scores}).
\item Gradient value $\NEXT_\alpha$ for next iteration of the algorithm (See Note \ref{next}).
\end{itemize}

\begin{enumerate}
\item Set $\CTR_\alpha$, called the {\em set of candidate target scores}, to the set of all scores $x$ where $f_{\bm{c}}(x) \geq \alpha$, and for all sufficiently small $\epsilon > 0$, $f_{\bm{c}}(x + \epsilon) \leq \alpha$. Sort it by descending score.

For the exact problem, $\CTR_\alpha$ is simply the set of all scores in $\bm{c}$'s support (See Note \ref{ctr}).

\item Initialize
\begin{itemize}
\item $\LTR$ (the {\em last trace}), to $\infty$
\item $\TRG_\alpha$ (the {\em set of target scores}),$\REI_\alpha$ (the {\em reinforcement set}) and $\COL_\alpha$ (the {\em collinear scores set}) to $\emptyset$
\item $\BDG_\alpha$ to $0$
\item $\NEXT_\alpha$ to $0$
\end{itemize}
\item While the highest $x \in \CTR_\alpha$ s.t. $x \leq \LTR$ exists
\begin{enumerate}
\item Add $x$ to $\TRG_\alpha$ (See Note \ref{dis}).
\item Find its trace $\TR_\alpha(x)$.

For the exact problem, find the highest $z \in \CTR_\alpha \cup \supp(\bm{a})$, $z < x$ such that $C_z^x \leq \alpha$. If there is no such $z$, $\TR_\alpha(x) = 0$. Otherwise $$\TR_\alpha(x) = x - \frac{(x - z) C_z^x}{\alpha}$$

\item For the exact problem only, update $\NEXT_\alpha$: Using $z$ from the previous step, if found, set $\NEXT_\alpha$ to $\max(C_z^x,\NEXT_\alpha)$.

\item Add $x$'s \em $\alpha$-segment, $[\TR_\alpha(x),x)$, unless it is empty, to $\REI_\alpha$ (See Note \ref{rei}).
\item If $\TR_\alpha(x) \in \CTR_\alpha$ ({\em collinear target score}) or $\TR_\alpha(x) \in \supp(\bm{a})$ ({\em collinear source score}), add $\TR_\alpha(x)$ to $\COL_\alpha$ (See Note \ref{col}).
\item Reinforce all entries in $\bm{a}$ whose scores are in $(\TR_\alpha(x),x)$ to $x$.
\item Add the budget used by the previous step, i.e., $$\frac{1}{|\mathcal{A}|}\sum_{z \in \mathcal{A} \land r_z \in (\TR_\alpha(x),x)} (x - r_z)$$ to $\BDG_\alpha$ (See Note \ref{bdg}).
\item Set $\LTR$ to $\TR_\alpha(x)$.
\end{enumerate}

\item For the statistical problem only, set $\NEXT_\alpha$ to $0$ if $\REI_\alpha$ is a single segment $[0, z)$ for which $F_{\bm{c}}(z) = 1$, and to $\alpha$ otherwise (See Note \ref{next}).
\label{this-item}
\end{enumerate}
\caption{Basic Algorithm}
	\label{basic}
\end{algorithm}

\begin{enumerate}
\item The parameter $\alpha$, in the context of p.d.f. $f_{\bm{c}}$, signifies a threshold density, while in the context of c.d.f. $F_{\bm{c}}$, signifies a gradient, i.e., the slope of a segment connecting two points on the c.d.f., or, equivalently, the average density between their scores/$x$-values.
\label{alpha}

\item
\label{next}
In the exact problem, $\BDG_\alpha$ is a step function of $\alpha$, due to Note \ref{bdg}. $\NEXT_\alpha$ is the {\em next-lower} gradient $\alpha$ at which $\BDG_\alpha$ changes. This is the gradient where the chord lines of the target scores, sweeping clockwise (lower $\alpha$) would meet a new entry (supported or complement) score. In Figures \ref{solution} and \ref{collinear}, this is the slope from target score $80$ to $120$, i.e., $\NEXT_\alpha = C_{80}^{120}$. The chord is marked in Figure \ref{solution} by a dotted green line. When no such entry exists, we set $\NEXT_\alpha = 0$.

In the statistical problem $\NEXT_\alpha = 0$ indicates that the entire complement c.d.f. is in the reinforcement set. This can be true only if $\bm{c}$ has bounded support.

In both versions of the problem, $\NEXT_\alpha = 0$ indicates that the current $\alpha$ uses the maximum possible budget: All supported entries outrank all complement entries.

\item
\label{ctr}
The scores in $\CTR_\alpha$ are characterized by the fact that the chord line of gradient $\alpha$ at $x$ is entirely on or above the complement's c.d.f. for a neighborhood surrounding {\em both} sides of $x$ (see Figure \ref{solution}).

$\CTR_\alpha$ includes
\begin{itemize}
\item All solutions $x$ of $f_{\bm{c}}(x) = \alpha$ where the density declines right of $x$. Equivalently, where the complement's c.d.f. tangent at $x$ is $\alpha$ and the c.d.f. is concave at $x$.
\item All scores $x$ where $\bm{c}$ has an atom.
\item Scores $x$ where $f_{\bm{c}}(x)$ has a non-atom discontinuity, with the left-density $\geq \alpha$ and the right-density $\leq \alpha$.
\end{itemize}

The set of candidate target scores is finite, {\em except} when the complement distribution is uniform with density $\alpha$ in all or part of its support. In this exception, the $\alpha$-segments are all empty, so $\TRG_\alpha$ is always finite.

For the  exact problem, the complement's c.d.f. is a step function, and the density is $0$ except at the steps. By the above definition, $\CTR_\alpha$ is the set of scores of all steps $x$, and furthermore, this set does not depend on $\alpha$ (but the set $\TRG_\alpha \subseteq \CTR_\alpha$ of target scores will, in general, depend on $\alpha$).
\item
\label{dis}
Since all chord lines have the same slope $\alpha$, they do not intersect. Therefore, every pair of $\alpha$-segments are either disjoint or one is wholly contained in the other. Since
$\TRG_\alpha$ includes only candidate target scores whose $\alpha$-segments are mutually disjoint, we can skip candidate target scores that are larger than the last target score's trace ($\LTR$), since their $\alpha$-segment must be wholly contained in the last $\alpha$-segment.
\item
\label{col}
For a collinear target score $z$, which is a trace of another target score $x$, the chords at $x$ and $z$ are necessarily collinear, and the endpoint of $z$'s $\alpha$-segment is the starting point of $x$'s $\alpha$-segment.

This is illustrated in Figure \ref{collinear}, where gradient $\alpha$ is such that the chord lines of scores $35$ and $80$ are collinear. 

On the other hand, in Figures \ref{solution} and \ref{collinear}, no supported scores coincide with a chord line trace, so they have no collinear source scores.

The presence of collinear scores, and certainly many of them, for any given value of $\alpha$, is exceptional. It requires the coincidence of, for example, a score in $\supp(\bm{a})$ coinciding with a trace of an atom in $\bm{c}$, or of chord gradients between two $\bm{c}$ atoms being equal to $\alpha$. There is a finite number of values of $\alpha$ with any collinear scores.
\item
\label{bdg}
The total budget used by Algorithm \ref{basic} is \footnote{Here and later $\HIGH(x)$ is shorthand for $\HIGH_{\REI_\alpha}(x)$. See Definition \ref{segments}.}
\s\begin{align*}
\BDG_\alpha &:= \int_{x \in \REI_\alpha \setminus \COL_\alpha} [\HIGH(x) - x]  f_{\bm{a}}(x) dx \\
& = \frac{1}{|\mathcal{A}|}\sum_{z \in \mathcal{A} \land r_z \in \REI_\alpha \setminus \COL_\alpha} [\HIGH(r_z) - r_z]
\end{align*}\n

\item
\label{rei}
The resulting $\alpha$-reinforcement set contains all reinforced scores and is \s$$\REI_\alpha := \bigcup_{x \in \CTR_\alpha} [\TR_\alpha(x), x) = \bigcup_{x \in \TRG_\alpha} [\TR_\alpha(x), x)$$\n

\end{enumerate}

\subsection{Iterative Algorithm}

The Iterative Algorithm, detailed in Algorithm \ref{supplementary}, applies the Basic Algorithm repeatedly using binary search to find the optimal solution for budget $p_A$, relying on the role played by the collinear scores, and on the monotonicity of $\BDG_\alpha$, properties that are stated and proved in Section \ref{analysis} below. The reader is referred to Section \ref{from-to} as to the method used by the algorithm and its justification.

\begin{algorithm}[tb]
Given an integral distribution $\bm{a}$, a complement distribution $\bm{c}$, budget $p_A$, and a sought accuracy $\epsilon \geq 0$ the algorithm outputs an integral reinforced distribution $\bm{A}$ in solution of Problem \ref{single-principal}.

For the statistical problem, set $\epsilon > 0$. For the exact problem, selecting $\epsilon = 0$ will yield the (non-approximate) optimal solution.

\begin{enumerate}
\item Set $\underline{\alpha}$ to an arbitrary value, and calculate Algorithm \ref{basic} for $\alpha = \underline{\alpha}$.
\item While $\BDG_{\underline{\alpha}} < p_A$ and $\NEXT_{\underline{\alpha}} > 0$
\begin{enumerate}
\item set $\underline{\alpha} \leftarrow \underline{\alpha}/2$
\item calculate Algorithm \ref{basic} for $\alpha = \underline{\alpha}$.
\end{enumerate}
\item If $\BDG_{\underline{\alpha}} < p_A$ (meaning that $\NEXT_{\underline{\alpha}} = 0$), the last solution is optimal, with all entries in $\mathcal{A}$ outranking all entries in $\mathcal{C}$. Exit.
\item Set $\overline{\alpha}$ to an arbitrary value, and calculate Algorithm \ref{basic} for $\alpha = \overline{\alpha}$.
\item While $\BDG_{\overline{\alpha}} \geq p_A$
\begin{enumerate}
\item Set $\overline{\alpha} \leftarrow 2\overline{\alpha}$
\item Calculate Algorithm \ref{basic} for $\alpha = \overline{\alpha}$.
\end{enumerate}
\item Binary search: While $\epsilon < \overline{\alpha} - \underline{\alpha}$
\begin{enumerate}
\item Set $\alpha = (\overline{\alpha} + \underline{\alpha})/2$, and calculate Algorithm \ref{basic} for $\alpha$.
\item If $\BDG_\alpha < p_A$ set $\overline{\alpha} \leftarrow \NEXT_\alpha$, else $\underline{\alpha} \leftarrow \alpha$ 
\end{enumerate}
\item Set $\alpha = \overline{\alpha}$. Recall Algorithm \ref{basic}'s calculated results for $\alpha$.
\label{end-binary}
\item If $|\COL_\alpha| = 0$
\label{col-0}
\begin{enumerate}
\item The last solution is the optimal one.
\end{enumerate}
\item Else if $|\COL_\alpha| = 1$
\label{col-1}
\begin{enumerate}
\item Let $y$ be the only collinear score, i.e., $\COL_\alpha = \{y\}$.
\item Set $K := \lfloor |\mathcal{A}| \frac{p_A - \BDG_\alpha}{\HIGH(y) - y} \rfloor$
\item Reinforce $K$ entries with score $y$ in $\bm{A}$ (from Algorithm \ref{basic}) to $\HIGH(y)$, i.e., subtract $K/|\mathcal{A}|$ from the atom currently at $y$ and add to, or create, the atom at $\HIGH(y)$.
\item The resulting solution, whose budget is $\BDG_\alpha + \frac{K}{|\mathcal{A}|}[\HIGH(y) - y]$, is the optimal one.
\end{enumerate}
\item Else ($|\COL_\alpha| > 1$)
\label{col>1}
\begin{enumerate}
\item Use any textbook solution of the bounded knapsack problem, e.g. \cite{martello1990knapsack}, with knapsack of $p_A - \BDG_\alpha$, values of $\HIGH(y) - y$ for each $y \in \COL_\alpha$, each with corresponding count $|\mathcal{A}| [F_{\bm{A}}(y) - \lim_{\epsilon \to 0} F_{\bm{A}}(y - \epsilon)]$ (= the number of supported entries with score $y$).
\end{enumerate}
\end{enumerate}

\caption{Iterative Algorithm}
\label{supplementary}
\end{algorithm}

\section{Analysis and Proof of Method}
\label{analysis}

\subsection{Analysis of the Basic Algorithm}

In this section, we demonstrate the correctness of the algorithms given in Section \ref{algorithm}.

\begin{definition}
If $\bm{A}$ is a solution of Problem \ref{single-principal}, define $\mathcal{R}(\bm{A}) := \{x > 0 | F_{\bm{a}}(x) > F_{\bm{A}}(x)\}$, i.e., the set of scores that were reinforced by the solution. Since the distributions are piecewise-continuous, $\mathcal{R}(\bm{A})$ is a union of segments. (E.g., in Figure \ref{reinforced}, $\mathcal{R}(\bm{A})$ consists of the $x$-values of the green areas).
\end{definition}

\begin{proposition}
\label{max}
Let $\bm{A}$ be a maximal solution of Problem \ref{single-principal}. Let $x$ be a target score and $x' \in \supp(\bm{A})$. If $y, y'$ are scores with $y' > x'$ and $\LOW_{\mathcal{R}(\bm{A})}(x) \leq y < x$, then $C_x^y \geq C_{x'}^{y'}$.
\end{proposition}

\begin{proof}
Consider an alternate distribution $\bm{A'}$, as follows: Select part of the reinforced density at $x$, say $\epsilon > 0$.\footnote{If changing the density at a single point, being infinitesimal, is considered a problem, envision the same procedure for a neighborhood $[x - \delta, x + \delta]$, with $\delta > 0$. Then let $\delta \to 0$.} Decrease that reinforced part by $x - y$ to $y$. Let $\epsilon' := \epsilon \frac{x - y}{y' - x'}$. Choose $\epsilon$ small enough so that $\epsilon < \inf_{z \in (y,x]} \big\{F_{\bm{a}}(z) - F_{\bm{A}}(z)\big\}$ and $\epsilon' < f_{\bm{A}}(x')$. Correspondingly, reinforce an $\epsilon'$ of the density at $x'$ to $y'$.

The choices $y \geq \LOW_{\mathcal{R}(\bm{A})}(x)$ and $\epsilon < \inf_{z \in (y,x]} \big\{F_{\bm{a}}(z) - F_{\bm{A}}(z)\big\}$ guarantee that the change cancels only part of the reinforcement from $\bm{a}$ to $\bm{A}$, so $\bm{A'}$ still stochastically dominates $\bm{a}$. Since $\epsilon (x - y) = \epsilon' (x' - y')$, we introduced a new reinforcement that compensates for the cancelled one, so $\bm{A'}$ has the same expectation as $\bm{A}$.

If $\bm{A}$ is an optimal solution of Problem \ref{single-principal}, then for every solution $\bm{A'}$ \s$$U(\bm{A}, \bm{c}) \geq U(\bm{A'}, \bm{c})$$\n So, by \eqref{U_c}
\s\begin{align}
\label{maximal}
 \int\limits_0^\infty \big[f_{\bm{A}}(y) - f_{\bm{A'}}(y)\big] F_{\bm{c}}(y) dy \geq 0
\end{align}\n

The difference between the densities is
\s\begin{align*}
f_{\bm{A}}(z) - f_{\bm{A'}}(z) = \left\{  \begin{array}{ll}
\epsilon &  z = x\\
\epsilon' & z = x' \\
-\epsilon & z = y \\
-\epsilon' & z = y'  \\
0 & $otherwise$ \\
\end{array} \right.
\end{align*}\n

Substituting the above in \eqref{maximal}
\s\begin{align}
\label{m1}
\epsilon F_{\bm{c}}(x) - \epsilon F_{\bm{c}}(y) + \epsilon' F_{\bm{c}}(x') - \epsilon' F_{\bm{c}}(y') \geq 0
\end{align}\n

Substituting $\epsilon' := \epsilon \frac{x - y}{y' - x'}$ in \eqref{m1}
\s\begin{align*}
&\epsilon \Big[F_{\bm{c}}(x) -  F_{\bm{c}}(y)\Big] \geq \epsilon \frac{x - y}{x' - y'} \Big[F_{\bm{c}}(x') - F_{\bm{c}}(y')\Big] \\
& \Rightarrow C_x^y \geq C_{x'}^{y'}
\end{align*}\n

\end{proof}

The following proposition shows that there is a dichotomy between two types of chords from a target score: All chords to the left (lower score) are steeper than all chords to the right, as well as to scores supported in $\bm{A}$. When $F_{\bm{c}}$ is differentiable at the target score, the dichotomy's boundary is the tangent slope at the target score.

\begin{proposition}
\label{split}
If $\bm{A}$ is a maximal solution of Problem \ref{single-principal}, then for every target score $x$, score $y \in [\LOW_{\mathcal{R}(\bm{A})}(x), x)$ and $x' > x$, or $x'$ in $supp(\bm{A})$,
\s$$C_x^y \geq C_x^{x'}$$\n and if $F_{\bm{c}}$ is differentiable at $x$, then \s$$C_x^y \geq f_{\bm{c}}(x) \geq C_x^{x'}$$\n
\end{proposition}

\begin{proof}
If $x' < x$, then by Proposition \ref{max} (substitute $x$ for $y'$) $C_x^y \geq C_x^{x'}$, while if $x' > x$, again by Proposition \ref{max} (substitute $x$ for $x'$ and $x'$ for $y'$) $C_x^y \geq C_x^{x'}$. In either case, $y$ and $x'$ may be as close to $x$ as desired. Therefore, if $F_{\bm{c}}$ is differentiable at $x$, letting $x' \to x$ we get $C_x^y \geq f_{\bm{c}}(x)$, and letting $y \to x$ we get $f_{\bm{c}}(x) \geq C_x^{x'}$.
\end{proof}

However, since a score may belong to {\em both} types, i.e., be both left of a target score, {\em and} belong to $\supp(\bm{A})$, we immediately conclude from Proposition \ref{split}:
\begin{corollary}
\label{colinear}
If $\bm{A}$ is a maximal solution of Problem \ref{single-principal} and $x$ is a target score,
all scores that are in both $\supp(\bm{A})$ and in $[\LOW_{\mathcal{R}(\bm{A})}(x), x)$ have the same gradient to $x$, so their chords are all collinear.

Furthermore, if $F_{\bm{c}}$ is differentiable at $x$, all such gradients are equal to $f_{\bm{c}}(x)$, and collinear with the tangent at $x$.
\end{corollary}

We can now characterize the optimal solution.

\begin{theorem}
\label{chord-line}
Let $\bm{A}$ be a maximal solution of Problem \ref{single-principal}. Then there is a positive real $\alpha$ (including $\alpha = \infty$), such that the chords and tangents of $F_{\bm{c}}$ satisfy:

For every reinforced segment $R \subseteq \mathcal{R}(\bm{A})$, $y := \HIGH(R)$ is a target score in $\TRG_\alpha$. $y$'s chord line, the line $F_{\bm{c}}(y) + \alpha (x - y)$ (Definition \ref{traces}), satisfies
\begin{enumerate}
\item \label{i1} $\TR_\alpha(y) \leq \LOW(R)$; I.e. $R$ is a subset of $y$'s $\alpha$-segment.
\item \label{i2} The c.d.f. of $\bm{c}$ in $R$ and everywhere $\geq y$ is entirely below or on the chord line.
\item \label{i3} If $F_{\bm{c}}$ is left-(right-)differentiable at $y$, its left (right) tangent is $\alpha$.
\item \label{i4} For every $x \in R$, $x \in \supp(\bm{A})$ implies $C_x^y = \alpha$, i.e., any target scores other than $y$, and any non-reinforced scores in the reinforced segment are on the chord line.
\item \label{i5} All scores in $R$ that are in $\supp(\bm{a})$, except possibly scores that are on the chord line, are reinforced to $y$. Those on the chord line are either not reinforced, or are reinforced to another score (including $y$) on the chord line.
\end{enumerate}
\end{theorem}

\begin{proof}
Item \ref{i1} follows from Proposition \ref{max}, items \ref{i2} \& \ref{i3}, from Proposition \ref{split}. Items \ref{i4} \& \ref{i5} follow from Corollary \ref{colinear}.
\end{proof}

Given a threshold density $\alpha$, Theorem \ref{chord-line} almost completely characterizes $\bm{A}$ after an optimal reinforcement: It shows that entries in $\mathcal{A}$ whose score falls in one of the $\alpha$-segments of a target score are reinforced to meet a target score in that segment. The only open question it leaves is whether that target score is the segment's high endpoint, or some other score that is on the chord line (this question, when relevant, is resolved by Algorithm \ref{supplementary}).

This proves the correctness of Algorithm \ref{basic}.
\begin{corollary}
Given $\alpha$, Algorithm \ref{basic} constructs the minimum-budget solution that complies with Theorem \ref{chord-line}.
\end{corollary}

 \subsection{Analysis of the Iterative Algorithm}
\label{from-to}

The Iterative Algorithm applies the Basic Algorithm repeatedly using numeric methods (binary search) to find the optimal (or $\epsilon$-optimal, for the statistical version) solution for budget $p_A$, relying on the role played by the collinear scores, as described in Theorem \ref{chord-line}, and on the monotonicity of $\BDG_\alpha$  and of $\REI_\alpha$, stated in the following proposition.

\begin{proposition}
\label{p_alpha}
$\BDG_\alpha$ is non-increasing in $\alpha$, and $\REI_\alpha \subseteq \REI_{\alpha'}$ for $\alpha > \alpha'$.
\end{proposition}

\begin{proof}
First we prove the following claim: Let $\alpha' < \alpha$. Then for every $y \in \CTR_\alpha$ there exists $y' \in \CTR_{\alpha'}$ such that $y' \geq y$ and $\TR_\alpha(y) \geq \TR_{\alpha'}(y')$. Proof: Let $y'$ be the smallest $y' \geq y$ such that for every small enough $\epsilon > 0$, $f_{\bm{c}}(y' + \epsilon) \leq \alpha'$. Such a $y'$ must exist since $\lim_{x \to \infty} f_{\bm{c}}(x) = 0$. Now, by the definition of the trace (Definition \ref{traces}), $\TR_\alpha(y)$ is the {\em highest} score that is $< y$ such that the average density of $\bm{c}$ in $[\TR_\alpha(y), y)$ is $\leq \alpha$. Similarly $\TR_{\alpha'}(y')$ is the {\em highest} score that is $< y'$ such that the average density of $\bm{c}$ in $[\TR_{\alpha'}(y'), y')$ is $\leq \alpha'$. From the way $y'$ was selected, the density of $\bm{c}$ everywhere in $[y,y)'$ is $> \alpha'$. Therefore, the average density in $[\TR_{\alpha'}(y'), y')$ is $\leq \alpha' < \alpha$. Therefore, we must have $\TR_\alpha(y) \geq \TR_{\alpha'}(y')$. This proves the claim.

It follows that every reinforced segment for every target score $y \in \CTR_\alpha$ is contained in a reinforced segment of some target score $y' \in \CTR_{\alpha'}$. Hence, Algorithm \ref{basic} for $\alpha'$ reinforces every score in $\bm{a}$ that Algorithm \ref{basic} for $\alpha$ does, and to a higher or equal target score. Consequently, its budget is at least as large, i.e., $\BDG_{\alpha'} \geq \BDG_\alpha$.
\end{proof}

The monotonicity of $\BDG_\alpha$ allows an $\epsilon$-optimal $\alpha$ to be found by binary search. Moreover, for the exact problem, the optimal solution can be found: $\BDG_\alpha$ is a non-increasing step function of $\alpha$, and the Basic Algorithm calculates the lower end of the step whose budget is $\BDG_\alpha$, $\NEXT_\alpha$. Whenever binary search halves the search interval, and the middle $\alpha$ needs a budget which is still short of the target $p_A$, $\NEXT_\alpha$, rather than the middle $\alpha$ becomes the lower end of the search interval. In combination with the binary search, this guarantees that the search will find the gradient $\alpha$ for the required budget $p_A$.

Theorem \ref{chord-line} allows for optimal solutions that are not constructed by the Basic Algorithm, where scores are reinforced to scores on the chord line that are {\em not} nearest. The solution remains optimal (though not for the same budget) if any entry that was reinforced to (or was originally at) a collinear score ($\in \COL_\alpha$) is promoted to the {\em next-higher} target score. 

Furthermore, if gradient $\alpha$ is slightly decreased to $\alpha - \epsilon$, for sufficiently small $\epsilon > 0$, the collinear scores disappear, and all entries that were reinforced by Algorithm \ref{basic} to a collinear score $y$ are now reinforced to the next target score that is not in $\in \COL_\alpha$. Mark it $\HIGHER(y)$. 
E.g., in Figure \ref{collinear}, scores $35$ and $80$ are collinear, so $\HIGHER(x) = 80$, for every score in $[\TR_\alpha(80), 80)$. We have a discontinuity in $\BDG_\alpha$ of size
\s\begin{align*}
\lim_{\epsilon \to 0} &\BDG_{\alpha - \epsilon} - \BDG_\alpha = \sum_{y \in \COL_\alpha} [\HIGHER(y) - y] [F_{\bm{a}}(y) - F_{\bm{a}}(\TR_\alpha(y))]
\end{align*}\n

It follows that for all budgets between $\BDG_\alpha$ and $\lim_{\epsilon \to 0} \BDG_{\alpha - \epsilon}$, the optimal solution has threshold density $\alpha$. This solution is implemented by first finding the basic solution with Algorithm \ref{basic}, and then ``promoting'' entries on collinear scores to a higher target score as much as the budget allows.

Since we seek an integral target distribution, in which an
integral number of entries are reinforced, and the promotion interval, i.e., the distance between target scores, is fixed, the problem amounts to finding an integral number for each of several given intervals. This adds, as closely as possible, to
the budget $p_A$, without exceeding it.
This is essentially a bounded knapsack problem, which, as is well-known, is NP-complete in $|\COL_\alpha|$, and pseudo-polynomial in $p_A$.

These considerations are handled by steps \ref{col-0} to \ref{col>1} of Algorithm \ref{supplementary}, according to the number of collinear scores $|\COL_\alpha|$ at the end of the binary search. If there are none, there is nothing to do, and the step \ref{end-binary} result is the solution.
When $|\COL_\alpha| = 1$, the knapsack problem is trivial, with only one kind of object. We find the number of entries to promote by dividing the knapsack size (i.e., the residual budget $p_A - \BDG_\alpha$) by the size of the object (the distance $\HIGHER(y) - y$, where $y$ is the sole collinear score), as done in Algorithm \ref{supplementary}, step \ref{col-1}.

When $|\COL_\alpha| > 1$, one may use any textbook solution of the bounded knapsack problem, e.g. \cite{martello1990knapsack}. While this problem is NP-complete in the number of collinear scores (but pseudo-polynomial in the residual budget), it is safe to assume that $|\COL_\alpha|$ is bounded: Indeed as noted in Note \ref{col}, there is only a finite number of values of $\alpha$, a real variable, for which there are any collinear scores, so $|\COL_\alpha| \leq 1$ almost surely.

\subsection{Running Time}
\label{time}

In the exact problem, the algorithm's running time depends on the total number of entries ranked $n := |\mathcal{W}| = |\mathcal{A}| + |\mathcal{C}|$, and on the {\em range} and {\em resolution} of the entries $\mathcal{W}$.
\s\begin{align}
\label{range}&\RANGE(\mathcal{W}) := \max_{i,j \in \mathcal{W}} |r_i - r_j| \\
\label{res}
&\RES(\mathcal{W}) := \min_{i,j \in \mathcal{W}, r_i \neq r_j} |r_i - r_j|
\end{align}\n

The running time of the Basic Algorithm is linear, i.e., $O(n)$ (assuming the ranking system already ranks them by score, otherwise the running time, including sort, is $O(n \log n)$). This is clear from the fact that the number of candidate target scores $|\CTR_\alpha|$ is at most $|\mathcal{C}| < n$, and for each, at most two chord gradients from it are computed (as starting point and as ending point). The number of operations performed on each entry in $\mathcal{A}$ is at most one, if it is reinforced.

The Basic Algorithm is executed several times in the Iterative Algorithm, so to find the total running time, we need to cap the number of iterations. The Iterative Algorithm starts with $\overline{\alpha}$ and $\underline{\alpha}$ whose traces are in the range of scores $\RANGE(\mathcal{W})$, defined in \eqref{range}. Using binary search, this range is cut in half every iteration, and ultimately it may need to be small enough to contain a single score, i.e., it should be smaller than the resolution of scores $\RES(\mathcal{W})$, defined in \eqref{res}. The number of iterations required is \s$O\Big(\log \frac{\RANGE(\mathcal{W})}{\RES(\mathcal{W})}\Big)$\n, and the total running time (for bounded $|\COL_\alpha|$) is \s$$O\Big(n \log \frac{\RANGE(\mathcal{W})}{\RES(\mathcal{W})}\Big)$$\n

\subsection{Unimodal Distributions}

When $\bm{c}$ has a piecewise-differentiable and unimodal probability density, this leads to an easy solution: For every threshold $\alpha$ there is exactly one candidate target score, and no collinear target scores. There is at most one collinear source score, the trace of the only target score.

By {\em unimodal}, we mean a p.d.f. that has a single maximum at $M \geq 0$, and is strictly increasing for $x < M$ and decreasing for $x > M$, strictly so when the density is non-zero.

Most commonly-used continuous distributions of a positive random value are unimodal: Exponential, log-normal, Gamma and power law distributions. A notable exception is the uniform distribution.

\begin{proposition}
\label{unimodal}
Let $f_{\bm{c}}(x)$ have piecewise-differentiable and unimodal density.
Then for every value of $\alpha$, Algorithm \ref{basic} has a single target score $h_\alpha \geq M$, with trace $l_\alpha := \TR_\alpha(h_\alpha)$, a reinforcement set $\REI_\alpha = [l_\alpha,h_\alpha)$, and budget
\s\begin{align}
\label{P}
\BDG_\alpha = \frac{1}{|\mathcal{A}|}\sum_{z \in \mathcal{A} \land r_z \in (l_\alpha, h_\alpha)} (h_\alpha - r_z)
\end{align}\n

\begin{enumerate}
\item The unique maximal solution of Problem \ref{single-principal} is generated by Algorithm \ref{basic} at $\alpha$ for which $p_A - \BDG_\alpha$ is non-negative and minimal, followed by promoting $K = \lfloor |\mathcal{A}| \frac{\BDG_\alpha - p_A}{h_\alpha - l_\alpha}\rfloor$ entries whose score is $l_\alpha$ to $h_\alpha$, if at least K such entries exist, or as many as exist otherwise.

\item Furthermore, if $l_\alpha \notin \supp(\bm{a})$, the above solution exactly achieves the budget, i.e., $\BDG_\alpha = p_A$. \label{exactly}
\end{enumerate}
\end{proposition}

\begin{proof}
For every $\alpha \in (0, f_{\bm{c}}(M)]$, the equation $f_{\bm{c}}(x) = \alpha$ has one and only one solution where $f_{\bm{c}}(x + \epsilon) \leq \alpha$ for all sufficiently small $\epsilon > 0$. Call it $h_\alpha$. We necessarily have $h_\alpha \geq M$. By Definition \ref{traces}, $h_\alpha$ is the only target score, and there are no target collinear scores. If $\TR_\alpha(h_\alpha) \in \supp(\bm{a})$, it is the only collinear source score. So we have $|\COL_\alpha| \leq 1|$. The proposition describes the effect of Algorithm \ref{supplementary} when there is a single target score and $\alpha$-segment, as well as at most one co-linear score.

The second part of the proposition states that when the $\alpha$ with the closest budget to $p_A$ has no co-linear scores, its budget is exactly $p_A$. To prove this, we show that $\BDG_\alpha$ is continuous in a neighborhood of $\alpha$. Since it is also monotonic by Theorem \ref{p_alpha}, we must have $\BDG_\alpha = p_A$.

We show that $\BDG_\alpha$ is continuous at $\alpha$. Since $f_{\bm{c}}(x)$ is strictly decreasing at $h_\alpha$, $h_\alpha$ is continuous in $\alpha$. By Proposition \ref{p_alpha} \s$$\REI_{\alpha + \epsilon} \setminus \REI_\alpha = [l_{\alpha + \epsilon}, l_\alpha) \cup [h_\alpha, h_{\alpha + \epsilon})$$\n
Since $\bm{a}$ is discrete, there is an $\epsilon$-neighborhood of $\alpha$ for which $l_{\alpha+\epsilon} \notin \supp(\bm{a})$, so that $ [l_{\alpha + \epsilon}, l_\alpha) \cap \supp(\bm{a}) = \emptyset$. So by \eqref{P}
\s\begin{align*}
\BDG_{\alpha + \epsilon} - \BDG_\alpha = &\sum_{z \in \mathcal{A} \land r_z \in (l_\alpha, h_\alpha)} (h_{\alpha + \epsilon} - h_\alpha) + \\
& \sum_{z \in \mathcal{A} \land r_z \in (h_\alpha, h_{\alpha+\epsilon})}  (h_{\alpha + \epsilon} - r_z)
\end{align*}\n
\noindent which limits at $0$ as $\epsilon \to 0$. Therefore, $\BDG_\alpha$ is continuous in this range.
\end{proof}

\begin{figure}[tbp]
\centering
\includegraphics[height=0.24\textheight]{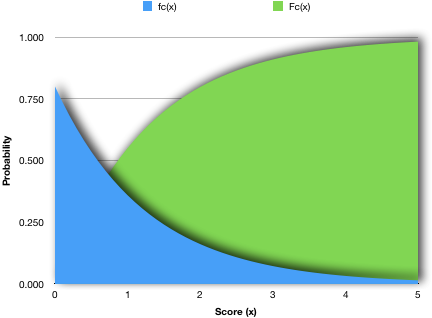}
\caption{Complement distribution $f_{\bm{c}}(x) = \lambda e^{-\lambda x}$ and $F_{\bm{c}}(x) = 1 - \lambda e^{-\lambda x}$ with $\lambda = 0.8$}
\label{exponential}
\end{figure}

\begin{example}
\label{decreasing}
Scores of $\bm{c}$ follow a monotonically-decreasing distribution $f_{\bm{c}}(x)$ (e.g., exponential distribution $f_{\bm{c}}(x) = \lambda e^{-\lambda x}$, see Figure \ref{exponential}) for positive $x$.

Let $n = |\mathcal{A}|$, and label the supported entries in ascending order of original scores $r_1 \leq  r_2 \leq \ldots \leq r_n$.

Solution: Since the mode $M$ is $0$, $\TR_\alpha(h_\alpha) = 0$ for every $\alpha$, and so Proposition \ref{unimodal}(\ref{exactly}) applies. Thus, we have an exactly optimal solution, when selecting a target score $h$, by \eqref{P}, for which
\s\begin{align*}
\frac{1}{n} \sum_{r_i < h} (h - r_i) = p_A
\end{align*}\n

This is achieved by the following algorithm:
\begin{enumerate}
\item Find the lowest $m$ s.t. $\sum_{i = 1}^m [r_{m+1} - r_i] > n p_A$, or set $m = n$ if there is no such $m$.
\item Set $h$ to $r_m + \frac{n p_A - \sum_{i = 1}^m [r_m - r_i]}{m}$.
\item Set $R_1 = \ldots = R_m = h$.
\end{enumerate}
\end{example}

\begin{example}
\label{lognormal}
Scores of $\bm{c}$ follow the log-normal distribution $LogNormal(0, 1)$, i.e., for positive $x$,
\s\begin{align*}
f_{\bm{c}}(x) &= \frac{1}{x \sqrt{2\pi}} e^{-\frac{(\ln x)^2}{2}} \\
F_{\bm{c}}(x) &= \frac{1}{2} + \frac{1}{2}\erf{\frac{\ln x}{\sqrt{2}}}
\end{align*}\n

Let $n = |\mathcal{A}|$, and label the supported entries in ascending order of original scores $r_1 \leq  r_2 \leq \ldots \leq r_n$.

Solution: The distribution is unimodal, with mode $M$ at $1 / e$. For every $\alpha \in (0,f_{\bm{c}}(M))$ there is therefore a single target score $h_\alpha > M$, at which $f_{\bm{c}}(h_\alpha) = \alpha$. Mark its trace by $l_\alpha := \TR_\alpha(h_\alpha)$.

If positive, $l_\alpha$ must satisfy $C_{l_\alpha}^{h_\alpha} = f_{\bm{c}}(h_\alpha)$, i.e., it solves
\s\begin{align}
\label{LH}
\frac{\erf{\frac{\ln h_\alpha}{\sqrt{2}}} - \erf{\frac{\ln l_\alpha}{\sqrt{2}}}}{2[h_\alpha - l_\alpha]} = \frac{1}{h_\alpha \sqrt{2\pi}} e^{-\frac{(\ln h_\alpha)^2}{2}}
\end{align}\n

For $h_\alpha > 2.232\ldots$, \eqref{LH} has no positive real solution, so $l_\alpha = 0$, and the solution is as given in Example \ref{decreasing}. When $1 / e < h_\alpha < 2.232\ldots$, $l_\alpha$ is the unique solution of \eqref{LH}, and we must select the unique $\alpha$ for which
\s\begin{align*}
\frac{1}{n} \sum_{l_\alpha < r_i < h_\alpha} (h_\alpha - r_i) \leq p_A \leq \frac{1}{n} \sum_{l_\alpha \leq r_i < h_\alpha} (h_\alpha - r_i)
\end{align*}\n

For the solution, apply Algorithm \ref{basic} for $\alpha$. If $l_\alpha \notin \supp(\bm{a})$, this yields the optimal solution, using the exact budget. If  $l_\alpha \in \supp(\bm{a})$, reinforce at most $K$ entries from $l_\alpha$ to $h_\alpha$, where
\s$$K := \Big\lfloor \frac{p_A - \frac{1}{n} \sum_{l_\alpha < r_i < h_\alpha} (h_\alpha - r_i)}{h_\alpha - l_\alpha} \Big\rfloor$$\n
\end{example}

\section{Discussion}
\label{discussion}

\subsection{Conclusion}

We studied the problem of how a principal, who owns or supports a set of ranked entries, can optimally allocate a budget to maximize their ranking.
We showed that, in general, the best ranking is achieved by equalizing the scores of several disjoint score ranges. We showed that there is a unique optimal reinforcement strategy, and provided an efficient, almost-surely linear algorithm implementing it.

\subsection{Non-Linear Utility}
By setting our target to optimize the average rank of the supported entries, our work presupposes that the principal's utility is {\em linear} in that metric, with results that it is often optimal to invest in the lowest-ranked ones. It may be argued that a realistic utility has a non-linear element, skewing the motivation in favour of higher-ranked entries. E.g., budget aside, promoting a YouTube video from rank 10 to 1 is more valuable than from rank 1000 to 991. Or, the promotion target may be to enter a top-hundred list, or likewise, which cannot be achieved by investing in low entries.

However, there is reason to believe our solution is close to optimal even with non-linear responses, unless the non-linearity, or the target, is contrived. Whether the target is to jump-start a trend (YouTube) or directly manipulate ranking (hotels), supporting the highest-ranked entries is likely to be poor strategy. Promoting a video from 1 million views to 1,010,000 views is likely a waste of money, while the budget could be spent to promote several novice videos from ``invisible'' status to ``almost-noticeable''. At writing, the difference between no. 1 and no. 2 on YouTube is 2 billion views, a huge budget, and enough to make a new video no. 32, or to give 2000 videos a million views. With hotels, one must remember that the principal's interest is in their aggregate welfare. When the target is making a top-hundred list, or similar, the target is either already achieved or unachievable, and in the rare cases that the strategy makes a difference, the problem is trivial.

\bibliographystyle{ACM-Reference-Format-Journals}
\bibliography{infomarkets}

\end{document}